\documentclass[runningheads,a4paper]{llncs}

\usepackage{amssymb}
\usepackage{graphicx}

\usepackage{url}
\urldef{\mailsa}\path|{alicia.canton, leonardo.fernandez, 
eugenia.rosado, mariajesus.vazquez}@upm.es|    
\newcommand{\keywords}[1]{\par\addvspace\baselineskip
\noindent\keywordname\enspace\ignorespaces#1}

\begin{document}

\mainmatter  

\title{Implicit equations of non-degenerate rational Bezier quadric triangles}

\titlerunning{Implicit equations of non-degenerate rational Bezier quadric triangle}

%
%
\author{A. Cant\'on \and L. Fern\'andez-Jambrina \and E. Rosado 
Mar\'\i a \and M.J. V\'azquez-Gallo%
}
\authorrunning{A. Cant\'on et al.}

\institute{Universidad Polit\'ecnica de Madrid,\\
28040-Madrid, Spain\\
\mailsa\\
\url{http://dcain.etsin.upm.es/~discreto/}}

%
%

\toctitle{Implicit equations of non-degenerate rational Bezier quadric triangle}
\tocauthor{L. Fern\'andez}
\maketitle

\begin{abstract}
In this paper we review the derivation of implicit equations for
non-degenerate quadric patches in rational B\'ezier triangular form.
These are the case of Steiner surfaces of degree two.  We derive the
bilinear forms for such quadrics in a coordinate-free fashion in terms
of their control net and their list of weights in a suitable form. 
Our construction relies on projective geometry and is grounded on the 
pencil of quadrics circumscribed to a tetrahedron formed by vertices 
of the control net and an additional point which is required for the 
Steiner surface to be a non-degenerate quadric.
\end{abstract}

\section{Introduction}

B\'ezier triangles \cite{triangle} are an alternative to tensor 
product patches as an extension of the B\'ezier formalism from curves to 
surfaces. In fact they were already present in De Casteljau's 
original work. Though they are not widely used as tensor product patches, 
they are useful in finite element methods and in gaming and 
animation, since the triangular geometry is more versatile for 
building surfaces and avoids the formation of singular points.

Quadrics are 
extensively used in engineering and therefore a usual requirement for 
a design formalism is that it may represent quadrics in an exact 
fashion. Quadric patches can be described as rational quadratic B\'ezier 
triangles, though not every rational quadratic B\'ezier 
triangle is a quadric patch. A characterisation can be found in 
\cite{hansford}. In general  rational quadratic B\'ezier 
triangles are quartic surfaces known as Steiner  surfaces. 
This family of surfaces includes ruled cubics and quadrics as 
subcases.

The relation between rational quadratic B\'ezier triangles and Steiner
surfaces has been studied since the very beginning of CAGD. In
\cite{sederberg} properties of Steiner surfaces are derived and they
are postulated as candidates for surface design. In \cite{lodha} a 
control polyhedron is used for representing quadric patches. The 
authors of \cite{dietz} define a generalised stereographic projection on the 
sphere to derive results for quadratic and biquadratic patches. 
 In \cite{coffman} 
 algebraic geometry is used for studying surfaces that can be parametrised 
quadratically. In \cite{degen} 
general B\'ezier triangles are studied as projections of Veronese 
surfaces and  the quadratic case is classified. 

In \cite{gudrunquadric} algebraic geometry methods are used to
determine whether a rational quadratic B\'ezier triangle is a quadric
patch and an algorithm is provided for classifying them.  A tool named
Weighted Radial Displacement is proposed for constructing B\'ezier
conics and quadrics in \cite{reyes-quadrics}.

In this paper we address the calculation of implicit equations for
non-degenerate quadrics in rational B\'ezier triangular form.  Our
goal is to find coordinate-free expressions that involve just the
control net and weights for the patch, using algebraic projective
geometry, as we did in \cite{conics}.  This is useful, for
instance, to compute geometric characteristics of the surfaces.

In Section~2 we review rational B\'ezier quadratic patches, introduce
notation and define a pencil of quadrics through the corners of the
control net of the patch and an additional point where the conics
located on the boundary of the patch meet.  In order to determine the
coefficients of the pencil of quadrics, in Section~3 we derive an
expression for the bilinear form of a conic circumscribed to a
triangle in terms of its control points and weights.  In Section~4 we
show that the data we have from each boundary conic of the
Steiner surface is compatible precisely if the surface is a quadric. 
In Section~5 we obtain the bilinear form for the Steiner quadric. 
Section~6 to several examples.

\section{Quadric Steiner surfaces}

We consider rational B\'ezier quadratic triangles,
\begin{equation}\label{steiner}c(u,v,w)=
\frac{\displaystyle\sum_{i+j+k=2}\frac{2!}{i!j!k!}\omega_{ijk}c_{ijk}u^iv^jw^k}
{\displaystyle\sum_{i+j+k=2}\frac{2!}{i!j!k!}\omega_{ijk}u^iv^jw^k},
\left.\begin{array}{c} u+v+w=1,\\\\ 
u,v,w\in[0,1],\end{array}\right.\end{equation} defined by 
its control points, $\{c_{002}, 
c_{011},c_{020},c_{101},c_{110},c_{200}\}$, and their respective weights, 
$\{\omega_{002}, 
\omega_{011},\omega_{020},\omega_{101},\omega_{110},\omega_{200}\}$, 
which are real numbers.

Such surface patches are bounded by three curves, defined 
respectively by the equations $u=0$, $v=0$, $w=0$. For instance, the 
arc at $u=0$ is parametrised by
\[c_{u}(v)=\frac{\displaystyle\sum_{j=0}^2
{2\choose j}\omega_{0j2-j}c_{0j2-j}v^j(1-v)^{2-j}}
{\displaystyle\sum_{j=0}^2{2\choose j}\omega_{0j2-j}v^j(1-v)^{2-j}},\quad
v\in[0,1],\]
and hence it is a conic arc with control polygon 
$\{c_{002},c_{011},c_{020}\}$ and weights  
$\{\omega_{002},\omega_{011},\omega_{020}\}$.

Similarly, the conic arc at $v=0$ has control polygon $\{c_{002},
c_{101}, c_{200}\}$ and weights $\{\omega_{002},\omega_{101},\omega_{200}\}$, 
whereas the control polygon of the one at $w=0$ is
$\{c_{020},c_{110},c_{200}\}$, with list of weights $\{\omega_{020},\omega_{110},\omega_{200}\}$.
 We assume from now on that these conics are non-degenerate.

The quadratic surface patch in (\ref{steiner}) is generically a 
quartic surface, named Steiner surface \cite{sederberg}, but in some 
particular cases it is a ruled cubic or a 
quadric. We are interested in the latter case due to the relevance of 
quadric surfaces. A characterisation of 
quadric Steiner surfaces is available in \cite{hansford}: 
\begin{figure}
\centering
\includegraphics[height=3.5cm]{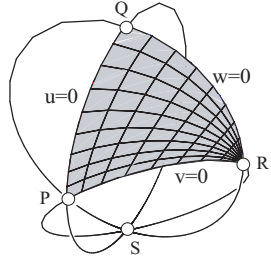}
\caption{Characterisation of quadric Steiner surfaces}
\label{characterisation}
\end{figure}
\begin{itemize}
    \item If the Steiner surface is a non-degenerate quadric, the
    three conic sections meet at a point $S$ and their tangents span a plane
    there (see Fig.~\ref{characterisation}).

    \item If the three conic sections meet at a point $S$ and their tangents
    span a plane there, the Steiner surface is a quadric.
\end{itemize}

The existence of point $S$ is useful for our purposes.  We label the
points at the corners of the surface patch as $P=c_{002}$,
$Q=c_{020}$, $R=c_{200}$.

\begin{figure}
\centering
\includegraphics[height=3cm]{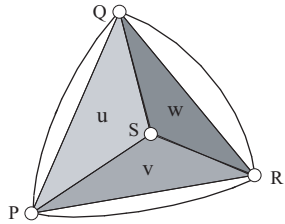}
\caption{Tetrahedron inscribed in a  quadric Steiner surface}
\label{tetrahedron}
\end{figure}

The three conic arcs  defined by $u=0$, $v=0$ and $w=0$ are 
respectively located at planes that we  denote $u$, $v$, $w$. We 
consider an additional plane $t$ through $P$, $Q$, $R$ (see 
Fig.~\ref{tetrahedron}). 

In order to simplify the notation, we also 
call $t,u,v,w$ the linear forms associated to the respective planes. 
Since they are defined up to a constant, we fix them by requiring
    \begin{equation}\label{normal}t(S)=u(R)=v(Q)=w(P)=1.\end{equation}

Since a quadric is determined by nine independent conditions, the
pencil of quadrics through $P$, $Q$, $R$, $S$ has five independent coefficients
\cite{kneebone}.  It is easy to check that the bilinear form $C$ for
such pencil in a coordinate-free fashion is
\begin{equation}\label{pencil}
C=\lambda_{tu}tu+\lambda_{tv}tv+\lambda_{tw}tw+\lambda_{uv}uv+
\lambda_{uw}uw+\lambda_{vw}vw,
\end{equation}
in terms of the linear forms for the planes containing the faces of 
the tetrahedron.


We have the bilinear form for the quadric except for the unknown
coefficients.  In the following section we determine the coefficients
$\lambda_{ij}$ by restricting $C$ to the planes
$u,v,w$.  Since the intersection of the quadric with such planes are
conics with known control polygons and weights, we determine the
coefficients up to proportionality factors.

\section{Bilinear forms for conic sections}

\begin{figure}
\centering
\includegraphics[height=3.5cm]{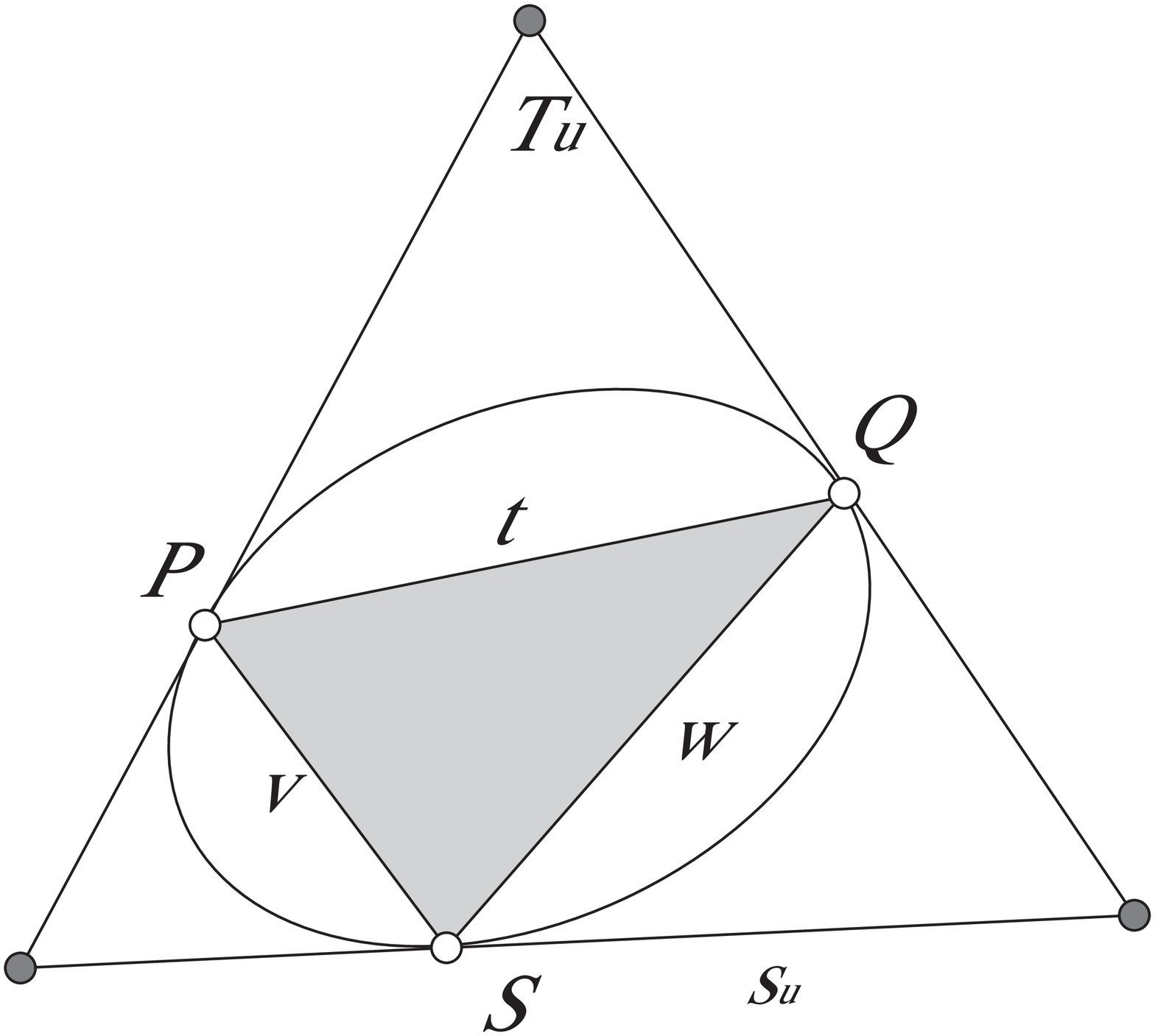}
\caption{Conic circumscribing a triangle}
\label{circumscribe}
\end{figure}

In order to determine the free coefficients of our pencil of 
quadrics, we need the bilinear forms for the conic sections of 
each of the faces of the tetrahedron. On Fig.~\ref{circumscribe} we have the 
conic on the face $u$ of the tetrahedron.

This conic has a bilinear form which is proportional to
\begin{equation}\label{conic}
C_{u}=\lambda_{tv}tv+\lambda_{tw}tw+\lambda_{vw}vw.\end{equation}

In this Section $t,v,w$ designate the straight lines which are the
intersections of the respective planes with the plane $u$, as well as
their linear forms.  That is, for simplicity in the notation in this
section, we identify $v$ with its restriction $v|_{u}$ on the plane
$u$ .

The conic on $u$ is determined by noticing that the polar line of 
the control point $T_{u}=c_{011}$ is $t$, since this is the line 
linking the points $P,Q$ where the tangent lines from $T_{u}$ meet 
the conic arc. That is, $C_{u}(T_{u},X)$ is proportional to $t(X)$ for all $X$ on the 
plane. For simplicity, we follow in most cases the following 
notation: the polar line of a point $A$ is the line $a$.

This conic arc from $P$ to $Q$ is parametrised by
\[c_{u}(t)=\frac{\omega_{002}P(1-t)^2+2\omega_{011}T_{u}t(1-t)+\omega_{020}Qt^2}
{\omega_{002}(1-t)^2+2\omega_{011}t(1-t)+\omega_{020}t^2},\qquad t\in[0,1],\]
but the weights are defined up to a M\"obius transformation of the 
interval $[0,1]$ onto itselt \cite{farin},
\[
    t(\tilde t)=\frac{\tilde t}{(1-\rho)\tilde t+\rho},\quad \tilde t\in[0,1],
\]
which produces a new list of weights for the same conic arc, 
\begin{equation}\label{weights}\tilde \omega_{002}=\rho^2\omega_{002},\quad \tilde \omega_{011}=\rho \omega_{011},\quad 
\tilde \omega_{020}=\omega_{020}.\end{equation}

We may use this degree of freedom to reparametrise the conic arc so that
\begin{equation}\label{scomb}S=\lim_{\tilde t\to\infty}c_{u}\left(t(\tilde t)\right)=\frac{\tilde \omega_{002}P-2\tilde \omega_{011}T_{u}+\tilde \omega_{020}Q}
{\tilde \omega_{002}-2\tilde \omega_{011}+\tilde \omega_{020}},\end{equation}
which has the advantage of writing the barycentric combination for 
$S$ in terms of the control polygon with coefficients that are simply 
the weights for the curve. Thus, 
\[T_{u}=\frac{\tilde 
\omega_{002}P+(2\tilde \omega_{011}-\tilde \omega_{002}-\tilde \omega_{020})S+\tilde 
\omega_{020}Q}{2\tilde \omega_{011}},\]
we can write the linear form for the polar line for $T_{u}$ in the
frame $\{P,S,Q\}$ with respect to the conic $C_{u}$ as
\begin{eqnarray*}
C_{u}(T_{u},X)
&\propto& \left(\lambda_{tv}\tilde \omega_{020}+\lambda_{tw}\tilde \omega_{002}\right)t(X)+
\left(\lambda_{tv}(2\tilde \omega_{011}-\tilde \omega_{002}-\tilde \omega_{020})+
\lambda_{vw}\tilde \omega_{002}\right)v(X)\\&+&
\left(\lambda_{tw}(2\tilde \omega_{011}-\tilde \omega_{002}-\tilde \omega_{020})+
\lambda_{vw}\tilde \omega_{020}\right)w(X),
\end{eqnarray*}
where the usual symbol $\propto$ means ``proportional to''.

Requiring that  $t$ be 
the polar line of $T_{u}$, 
we get the unknown coefficients of its bilinear form,
\begin{equation}C_{u}\propto\tilde \omega_{002}tv+\tilde \omega_{020}tw+(\tilde \omega_{002}-2\tilde 
\omega_{011}+\tilde \omega_{020})vw.\end{equation}

If $S$ were a point at infinity, instead of  (\ref{scomb}) we would need
\[S=\tilde \omega_{002}P-2\tilde \omega_{011}T_{u}+\tilde \omega_{020}Q \Rightarrow 
T_{u}=\frac{\tilde \omega_{002}P-S+\tilde \omega_{020}Q}{2\tilde \omega_{011}}
,\]
and we read the coefficients again imposing that $t$ is the polar 
line of $T_{u}$,
\begin{equation}C_{u}\propto\tilde \omega_{002}tv+\tilde \omega_{020}tw+vw.\end{equation}

We can summarise this result in the following lemma: 
\begin{lemma}The bilinear form for a conic circumscribed to a triangle $PQS$ with sides 
$t,v,w$ as in Fig.~\ref{circumscribe} is
\[\left\{
\begin{array}
{lcl}
\tilde \omega_{002}tv+\tilde \omega_{020}tw+(\tilde \omega_{002}-2\tilde 
\omega_{011}+\tilde \omega_{020})vw, &&\mathrm{if}\ \tilde \omega_{002}-2\tilde 
\omega_{011}+\tilde \omega_{020}\neq 0, \\[0.1cm]
\tilde \omega_{002}tv+\tilde \omega_{020}tw+vw,    
     && \mathrm{if}\ \tilde \omega_{002}-2\tilde 
\omega_{011}+\tilde \omega_{020}=0,\end{array}
\right.\]
where $S=\tilde \omega_{002}P-2\tilde \omega_{011}T_{u}+\tilde \omega_{020}Q$ up to a 
constant.\end{lemma}

The tangent line $s$ to the conic at $S$ is then $\tilde
\omega_{002}v+\tilde \omega_{020}w$, with tangent vector
$\vec{s}_{u}=\tilde \omega_{020}\overrightarrow{SP}- \tilde
\omega_{002}\overrightarrow{SQ}$.


\section{Reparametrising the quadric}

The latter theorem provides some of the unknown coefficients in
(\ref{pencil}) up to a constant.  Since we have made use of a special choice of weights
on $u$ to reach this result, we have to check that we can make it on 
the three boundary conics at a time in order to apply it to the whole
quadric. We try to reparametrise the three boundary conics as in 
(\ref{scomb}).



After reparametrising the conic on the plane $u$, the new list of 
weights is
\[\{\rho^2\omega_{002}, \rho \omega_{011},\omega_{020},\rho  \omega_{101},\omega_{110},\omega_{200}\},\]
for some constant $\rho$ and we obtain a tangent vector 
$\vec{s}_{u}=\omega_{020}\overrightarrow{SP}-
\rho^2 \omega_{002}\overrightarrow{SQ}$ to the conic at $S$.

If we reparametrise the conic arc from $R$ to $Q$ 
on the plane $w$, the list of weights changes again,
\[\{\rho^2\omega_{002}, \rho \omega_{011},\omega_{020},\sigma \rho  
\omega_{101},\sigma \omega_{110},\sigma^2 \omega_{200}\},\]
for some constant $\sigma$ and we get a new tangent vector 
$\vec{s}_{w}=\omega_{020}\overrightarrow{SR}-
\sigma^2 \omega_{200}\overrightarrow{SQ}$ to the conic at $S$.

Finally, if we needed to reparametrise the conic arc from $P$ to $R$ on the plane 
$v$, the list of weights would change to
\[\{\tau^2\rho^2\omega_{002}, \tau\rho \omega_{011},\omega_{020},\tau\sigma \rho  
\omega_{101},\sigma \omega_{110},\sigma^2 \omega_{200}\},\] for some constant $\tau$ and we 
would obtain another tangent vector $\vec{s}_{v}=\sigma^2\omega_{200}\overrightarrow{SP}-
\tau^2\rho^2 \omega_{002}\overrightarrow{SR}$ to the conic at $S$.

The last reparametrisation obviously spoils the previous ones, but we 
may check whether it is necessary or not.

If the Steiner patch is a non-degenerate quadric, the three tangent 
vectors are to lie on a plane \cite{hansford}. The determinant of these vectors,
\[\det(\vec{s}_{u},\vec{s}_{v},\vec{s}_{w})=
\left|\begin{array}{ccc} \omega_{020} & -\rho^2 \omega_{002} & 0  \\
\sigma^2\omega_{200} & 0 & -\tau^2\rho^2 \omega_{002}  \\
0 & -\sigma^2 \omega_{200} & \omega_{020}\end{array}\right|=
\rho^2\sigma^2\omega_{002}\omega_{020}\omega_{200}\left(1-\tau^2\right),\]tells us 
that they form a plane if and only if $\tau=1$, that is, the 
reparametrisations to locate $S$ at $t=\infty$ on the three conics 
are compatible. 


From now on we omit the tildes over the weights, assuming that we are 
using a set of weights with this property,
\begin{eqnarray}\label{baryS}S&=&
\frac{\omega_{002}c_{002}-2\omega_{011}c_{011}+ \omega_{020}c_{020}}{\omega_{002}-2\omega_{011}+ \omega_{020}}=
\frac{\omega_{002}c_{002}-2\omega_{101}c_{101}+\omega_{200}c_{200}}{\omega_{002}-2\omega_{101}+ \omega_{200}}
\nonumber\\&=&
\frac{\omega_{200}c_{200}-2\omega_{110}c_{110}+\omega_{020}c_{020}}{\omega_{200}-2\omega_{110}+ \omega_{020}},
\end{eqnarray}
if $S$ is a point. If it is a point at infinity, $S$ has in principle three different 
representatives for each conic,
\[S_{u}=\omega_{002}c_{002}-2\omega_{011}c_{011}+ \omega_{020}c_{020}, \quad
S_{v}=\omega_{002}c_{002}-2\omega_{101}c_{101}+\omega_{200}c_{200}, \]\[
S_{w}=\omega_{200}c_{200}-2\omega_{110}c_{110}+\omega_{020}c_{020},\]
which are parallel vectors. 
We write  the bilinear form 
for the conic on  $u$ as
\[\omega_{002}tv+ \omega_{020}tw+t(S_{u})vw,\]
to overcome this problem.

\section{Bilinear forms for Steiner quadrics}

If $S$ is a point, we have obtained  bilinear forms for 
the conics on  $u,v,w$ as
\[C_{u}\propto \omega_{002}tv+ \omega_{020}tw+( \omega_{002}-2 
\omega_{011}+ \omega_{020})vw,\]
\[C_{v}\propto \omega_{002}tu+ \omega_{200}tw+( \omega_{002}-2 
\omega_{101}+ \omega_{200})uw,\]
\[C_{w}\propto \omega_{020}tu+ \omega_{200}tv+( \omega_{020}-2 
\omega_{110}+ \omega_{200})uv,\]
and we  can fit all pieces of information in the bilinear form:
\begin{theorem}The bilinear form for a non-degenerate Steiner quadric 
patch, bounded by three non-degenerate conic arcs, with vertices of 
the control net $\{c_{002}, 
c_{011},c_{020},c_{101},c_{110},c_{200}\}$ and  weights  
$\{\omega_{002}, 
\omega_{011},\omega_{020},\omega_{101},\omega_{110},\omega_{200}\}$, fulfilling that the 
intersection $S$ of the  boundary conics is written as in 
(\ref{baryS}) is 
\begin{eqnarray*}C&=& \omega_{020} \omega_{002}tu+
    \omega_{002} \omega_{200}tv+ \omega_{200} \omega_{020}t w+
 \omega_{002}( \omega_{020}-2 \omega_{110}+ \omega_{200})uv\\&+&
 \omega_{200}( \omega_{002}-2 \omega_{011}+ \omega_{020})v w+
 \omega_{020}( \omega_{002}-2 \omega_{101}+ \omega_{200})uw,
\end{eqnarray*}
where $u$ is the linear form of the plane containing $c_{002},
c_{011},c_{020}$ which satisfies $u(c_{200})=1$, $v$ is the linear
form of the plane containing $c_{002}, c_{101},c_{200}$ which
satisfies $v(c_{020})=1$, $w$ is the linear form of the plane
containing $c_{020}, c_{110},c_{200}$ which satisfies $w(c_{002})=1$ 
and  $t$ is the linear form of the plane
containing $c_{002},c_{020}, c_{200}$ which satisfies $t(S)=1$.

If $S$ is a point at infinity, the bilinear form is just 
\begin{eqnarray*}C&=& \omega_{020} \omega_{002}tu+
    \omega_{002} \omega_{200}tv+ \omega_{200} \omega_{020}t w\\&+&
 \omega_{002}t(S_{w})uv+ \omega_{200}t(S_{u})v w+ \omega_{020}t(S_{v})uw.
\end{eqnarray*}
\end{theorem}

This result provides a procedure for computing a bilinear form 
 for a non-degenerate Steiner quadric 
patch in a coordinate-free fashion using just the vertices of the 
control net and their respective weights:

\begin{enumerate}
    \item  Compute the normalised linear forms for the planes $t,u,v,w$.

    \item Obtain $S$ as intersection of the planes $u,v,w$ and check
    if the patch belongs to a non-degenerate quadric.

    \item  Obtain an equivalent list of weights fulfilling 
    (\ref{baryS}).

    \item  Use Theorem~1 to obtain the bilinear form for the quadric 
    patch.

    \item  The implicit equation for the quadric patch is then $C(X,X)=0$.
\end{enumerate}

\section{Examples}

\begin{figure}
\centering
\includegraphics[height=3cm]{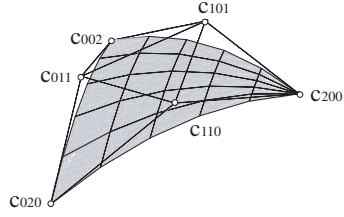}
\caption{Elliptic paraboloid}
\label{paraboloid}
\end{figure}

We use the previous results to compute implicit equations for several 
quadric patches:

\begin{example} Net: $\left[\begin{array}{c}
(0,0,0)\hspace{0.2cm}(1,0,1)\hspace{0.2cm}(2,0,0)\\
(0,1,1)\hspace{0.2cm}(1,1,1)\\ (0,2,0)\end{array}\right]$ and  
 weights: $\left[\begin{array}{c}1\hspace{0.2cm}1\hspace{0.2cm}1\\
1\hspace{0.2cm}1\\
1\end{array}\right]$ (Fig.~\ref{paraboloid}):\end{example}

The faces of the tetrahedron are the planes \[u: \frac{y}{2}=0,\quad v: \frac{x}{2}=0,\quad w: 
1-\frac{x+y}{2}=0, \quad t: -\frac{z}{2}=0.\]

The planes $u,v,w$ meet at the point at infinity 
\[S=(0,0,-2)=c_{002}-2c_{011}+c_{020}=
c_{002}-2c_{101}+c_{200}=c_{200}-2c_{110}+c_{020}.\]

The linear forms for the planes have been normalised according to 
(\ref{normal}). Hence the bilinear form for this surface is
\[C=tu+tv+t w+uv+v w+uw,\] and the implicit equation, in cartesian 
coordinates is
\[0=\frac{2x+2y-x^2-y^2-xy-2z}{4},\]
which corresponds to an elliptic paraboloid.

\begin{figure}
\centering
\includegraphics[height=4cm]{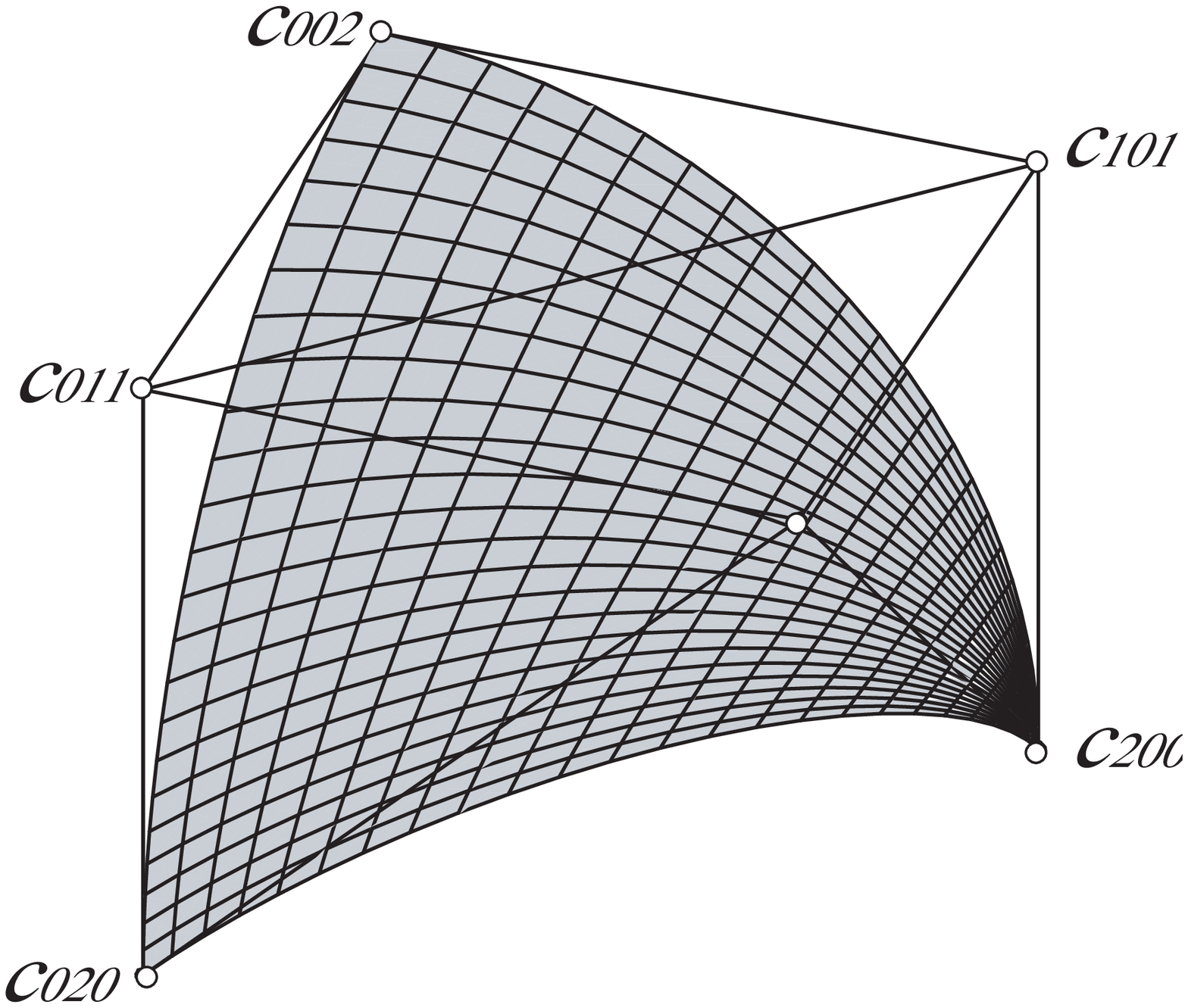}
\caption{Sphere}
\label{sphere}
\end{figure}

\begin{example} Net: 
$\left[\begin{array}{c}(0,0,1)\hspace{0.2cm}(1,0,1)\hspace{0.2cm}(1,0,0)\\(0,1,1)\hspace{0.2cm}(1,1,1)\\
(0,1,0)\end{array}\right]$ and weights: $\left[\begin{array}{c}1 \hspace{0.2cm} 1 \hspace{0.2cm} 2\\
1 \hspace{0.2cm} 1\\ 2\end{array}\right]$ (Fig.~\ref{sphere})\end{example}

The faces of the tetrahedron are the planes \[u: y=0,\quad v: x=0,\quad w: 
\frac{1-x-y+z}{2}=0, \quad t: \frac{1-x-y-z}{2}=0.\]

The planes $u,v,w$ meet at the point $S=(0,0,-1)$. The bilinear form 
for this surface is
\[C=2tu+2tv+ 4t w+2uv+2v w+2uw,\] and the implicit equation in cartesian coordinates is
\[0=1-x^2-y^2-z^2,\]
which corresponds to a sphere.

\begin{figure}
\centering
\includegraphics[height=2.5cm]{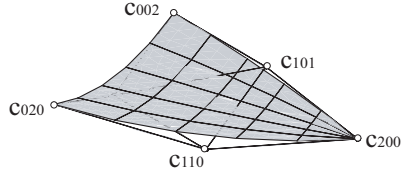}
\caption{Hyperbolic paraboloid}
\label{paraboloid1}
\end{figure}

\begin{example} Net: $\left[\begin{array}{c}(0,0,0)\hspace{0.2cm}(1,0,0)\hspace{0.2cm}(2,0,2)\\
(0,1/2,0)\hspace{0.2cm}(1,1/2,0)\\ (0,1,-1/2)\end{array}\right]$ and  
 weights: $\left[\begin{array}{c}1\hspace{0.2cm}1\hspace{0.2cm}1\\1\hspace{0.2cm}1\\ 
1\end{array}\right]$ (Fig.~\ref{paraboloid1}):\end{example}

The faces of the tetrahedron are the planes \[u: y=0,\quad v: \frac{x}{2}=0,\quad w: 
1-\frac{x}{2}-y=0, \quad t: \frac{2z-2x+y}{4}=0.\]

The planes $u,v,w$ meet at the point at infinity 
\[S=(0,0,2)=c_{002}-2c_{011}+c_{020}=-4(
c_{002}-2c_{101}+c_{200})=\frac{4}{3}(c_{200}-2c_{110}+c_{020}),\]
and with the choice of bilinear form for $t$ we have
\[t(S_{u})=1,\quad t(S_{v})=-\frac{1}{4},\quad t(S_{w})=\frac{3}{4}.\]

Hence the bilinear form for this surface is
\[C=tu+tv+t w+\frac{3}{4}uv+v w-\frac{1}{4}uw,\] and the implicit equation, in cartesian 
coordinates is
\[0=\frac{2z-x^2+y^2}{4},\]
which corresponds to a hyperbolic paraboloid.


%
%
%

\bibliographystyle{splncs03}
\bibliography{cagd}

\end{document}